\documentclass[
reprint,
prb,
floatfix,
]{revtex4-1}

\usepackage{CJK}
\usepackage{graphicx}
\usepackage{dcolumn}
\usepackage{bm}
\usepackage{hyperref}
\usepackage{xcolor}
\hypersetup{
    colorlinks=true,
    linkcolor=blue,
    filecolor=magenta,
    citecolor = blue,
    urlcolor=blue,
}
\usepackage{caption}
\usepackage{subcaption}

\graphicspath{{Pictures/}}
\begin{document}

\title{High resolution spectroscopy of thulium atoms implanted in solid noble gas crystals.}

\author{Vinod Gaire}
 \email{vgaire3@gatech.edu}
 \author{Mi Y Do}
 \email{mdo38@gatech.edu}
 \author{Yiting Pei}
 \email{ypei42@gatech.edu}
 \author{Anthony Semenova}
 \email{ksa6@gatech.edu}
\author{Colin V. Parker}%
 \email{cparker@gatech.edu}
\affiliation{%
 School of Physics, Georgia Institute of Technology, Atlanta, Georgia, 30332, USA\\
 }

\date{\today}
\pacs{Valid PACS appear here}

\begin{abstract}

Optically active defects in solid-state systems have many applications in quantum information and sensing. However, unlike free atoms, which have fixed optical transition frequencies, the inhomogeneous broadening of the transitions in solid-state environments limit their use as identical scatterers for such applications. Here we show that crystals of argon and neon prepared in a closed-cycle cryostat doped with thulium atoms at cryogenic temperatures are an exception.  High resolution absorption and emission spectroscopy show that the 1140 nm magnetic dipole transition is split into multiple components. The origin of this splitting is likely a combination of different classes of trapping sites, crystal field effects within each site, and hyperfine interactions. The individual lines have ensemble widths as small as $0.6\textrm{ GHz}$, which temperature dependence and pump-probe spectroscopy indicate is likely a homogeneous effect, suggesting inhomogeneity is well below the GHz scale.
\end{abstract}
\maketitle

\section{Introduction}
 With the advancement of quantum technologies, new sensing applications are emerging that offer new capabilities as well as the ability to probe new systems and environments. Various types of quantum sensors, including clocks \cite{kitching2011atomic,ushijima2015cryogenic,Camparo2007}, interferometers and gyroscopes \cite{Biedermann2017,Charriere2012,Mcguinness2012}, and magnetometers, \cite{Wickenbrock2013,Patton2014} are being developed from isolated atoms in vapor phase. However, building such sensors with isolated atoms requires a hermetically sealed environment, often within ultra high vacuum, and yields low atomic density. Alternatively, one can use solid-state platforms, including superconducting quantum interferometer devices (SQuIDs) \cite{Zotev_2007,MATLACHOV20041,Gatteschi2003,Sawicki2011} and color centers in diamond \cite{barry2020,lesik2019magnetic,childress2013,Acosta2009,Taylor2008}. However, such systems typically experience material inhomogeneity effects that lead to variation between fabricated devices or between individual color centers \cite{rogers_multiple_2014, Dam2019}. A third alternative is atoms embedded in solid noble gases, where the inertness of the host material limits inhomogeneous effects. Compared to vapor phase atoms, these so-called ``matrix-isolated'' systems offer higher density, can be deposited arbitrarily close to the surface of any desired substrate or device, and can be co-deposited with molecules of interest. We are interested in such systems with narrow linewidths for detection of the environment surrounding the target atoms. In the majority of matrix isolation work with atoms, however, alkali atoms have been used where the optical transitions suffer broadening at the THz level despite the inert host \cite{kanagin2013optical}.
 
 In our previous work, we observed that when thulium atoms are trapped in argon and neon crystals, the magnetic dipole transition between the fine structure levels of the ground state (${}^2F_{7/2}\leftrightarrow {}^2F_{5/2} $) was split into at least two components with linewidth less than 1 nm, far narrower than in alkali metals, and limited by the spectrometer itself. The transition was studied by excitation of higher energy levels with visible light and detection of fluorescence on the infrared transition near 1140 nm \cite{gaire2019}. Since the partially filled 4$f$ shell is submerged underneath the fully filled outer $5s$ and $5p$ shells, this transition is isolated from the external environment and was also observed to be narrow when trapped in solid and liquid helium \cite{ishikawa1997}. In this work, we perform both absorption and emission spectroscopy with far better resolution using laser-induced fluorescence at 1140 nm, revealing that this line in fact splits into many components with linewidths down to 0.6 GHz, likely due to a combination of multiple host trapping sites and crystal field effects at each site. Furthermore, we use the temperature dependence of the linewidth and the lack of spectral hole burning to argue that this linewidth is homogeneously broadened, suggesting MHz-scale population linewidths are possible in the solid state if this system is cooled further. If this prediction proves correct, the Tm:Ar system could contend for the least inhomogeneously-broadened solid-state optical emitter known. Even the currently obtained value of 0.6 GHz full-width half-maximum (FWHM) for an ensemble width is significantly below most other solid-state systems; for example, Eu$^{3+}$ in Y$_2$O$_3$ thin films (5.1 GHz at temperatures $<$ 100 mK \cite{Singh2020}), NV centers in diamond (5.6 GHz \cite{Dam2019}) and Yb$^{3+}$:YAG ( 3.6 GHz\cite{Bottger2016}). Comparison to solid-state platforms that can resolve individual color centers (such as SiV \cite{rogers_multiple_2014}) is less direct, but considering the distribution to be normal with a standard deviation of 300 MHz, we estimate an ensemble line would be Gaussian with FWHM of 0.7 GHz.\\
 
\section{Experimental Setup}
\begin{figure}
    \centering
    \includegraphics[width=\columnwidth]{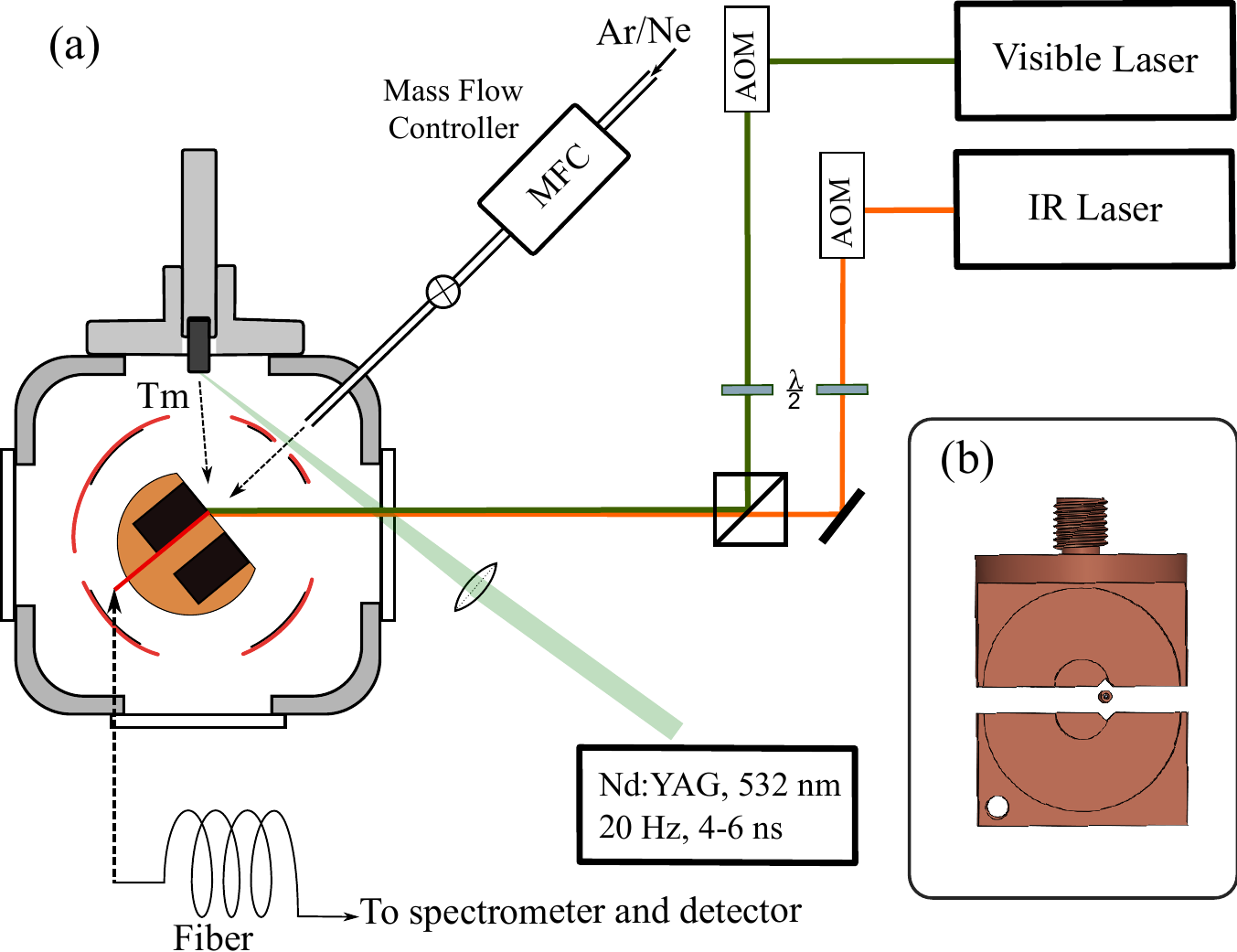}
    \caption{(a) Setup for spectroscopy of thulium atoms trapped in argon or neon crystals. (b) Mount for clamping the fiber tip, which is glued into a small copper tube. The circular groove is for holding an annular permanent magnet. }
    \label{fig:setup}
\end{figure}
A schematic of the cryostat chamber and the basic beam paths for the excitation lasers used in the matrix isolation experiment is shown in Fig. \ref{fig:setup}, similar to that used in \cite{gaire2019}. However, unlike in previous work, here the inert gas (argon or neon) is condensed directly on a multimode optical fiber tip (50 $\mu $m core diameter, NA 0.2). The thulium atoms are co-deposited with the matrix by ablation with a pulsed laser.  The fiber tip is glued with thermally conductive epoxy into a copper tube with outer diameter 1.6 mm. A purpose-built mount attached to the second stage of the cryostat, shown in the inset of  Fig. \ref{fig:setup}, holds the tip-tube assembly clamped with indium metal into a V-shaped groove. This mount also can hold an annular permanent magnet (dimensions: 25.4 mm O.D., 7.9 mm I.D., 6.4 mm thickness) if needed for doing experiments with magnetic field near the atoms.

\begin{figure}
    \centering
    \includegraphics[width=\columnwidth]{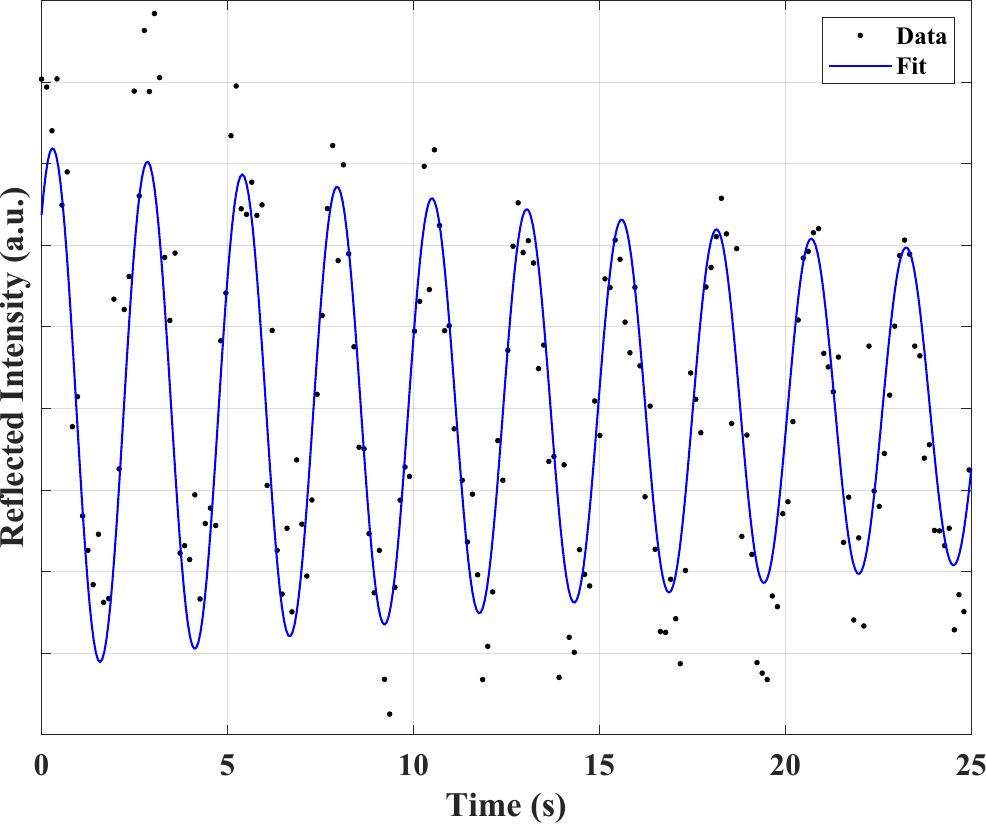}
    \caption{Oscillations in the monitor laser as the neon flow begins. The period of oscillation is used to monitor the growth rate of the sample.}
    \label{fig:thinfilm}
\end{figure}

In the beginning, when the setpoint temperature is reached, there is a period of 30-60 seconds when only undoped argon or neon is deposited. During this period, the deposition rate of the noble gas can be recorded \textit{in situ} using a diode laser. A diode laser illuminates the fiber at the other end of the sample, and the reflected beam spot is detected as it comes out from the fiber and separated from the incident light by a beam splitter. As the thickness of the crystal increases, the reflected light intensity undergoes Fabry-Perot oscillations as shown in Fig.\ref{fig:thinfilm}. The thickness (\textit{t})  of the sample is given by $2tn_{rg} = m\lambda$, where $n_{rg}$ is the refractive index of the rare gas used, and $\lambda $ the wavelength of monitoring laser. By fitting the data to a sinusoidal curve with an exponentially decaying term, the crystal growth is calculated and extrapolated for 30 minutes of sample growth to determine the total thickness of the sample. Using the period of Fabry-Perot oscillation, the deposition rate of the crystals calculated for our case is about 100-200 nm/s. 

After this initial period of undoped crystal growth, which also helps in protecting the fiber from becoming coated with thulium, the pulsed laser is turned on to dope the remaining growth with thulium atoms. The ablation is done for 10-30 minutes, so the total thickness of the crystal is $60-360\textrm{ }\mu\textrm{m}$. At the end of the growth, the temperature is 5 K at the cold head and 6.9 K on the lower part of the clamp. The sample may be hotter than either of these, as the cooling path to the sample runs through friction mounts, adhesive, optical fiber, and the rare gas matrix itself. We estimate the density of the thulium atoms is in parts per thousand.

\begin{figure}
    \centering
    \includegraphics[width=\columnwidth]{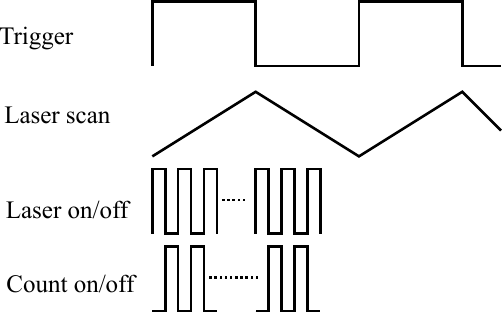}
    \caption{The timing sequence for the absorption spectroscopy.}
    \label{fig:timing}
\end{figure}

The sample is excited and the fluorescence signal is collected through the fiber and detected using an InGaAs single photon avalanche diode (SPAD). The experiment and data acquisition sequence are synchronized by an internal clock from a microcontroller. To perform absorption spectroscopy, a function generator sweeps the laser frequency by modulating the cavity length via a triangle wave voltage to the piezoelectric actuator. This scans the laser continuously over a mode-hop-free range of about 0.01-0.02 nm. While the laser frequency is being scanned, the laser is turned on/off by the TTL pulses generated by the microcontroller and amplified to get the required logic output for use with a radio frequency switch and the AOM. The light is on for a period of 4.9 ms, and the counting begins  200 $\mu s$ after the laser is turned off and continues for another 5.1 ms. A synchronized TTL signal triggers the microcontroller to begin acquisition at the start of each sweep. For detailed scans, we sweep every 22 s; for faster scans, we sweep every 2 s. The acquisition is done while the excitation light is switched off, and the timing sequence for absorption spectroscopy is shown in Fig. \ref{fig:timing}. One scan is completed after a set number of pulses, which occurs slightly before the next trigger is received. The acquired data is averaged for multiple such scans. Usually, only the second half of the acquired data is used, thus discarding the data where the wavelength of the laser is decreasing. Intensity for spectroscopy is typically a few $\textrm{W}/\textrm{cm}^2$.

Emission spectroscopy involves a similar sequence synchronized to the motion of a linear translation stage, forming a Michelson interferometer and allowing us to operate as a Fourier transform spectrometer as described in \cite{gaire2022_ftir}.

\begin{figure}
    \centering
    \includegraphics[width=\columnwidth]{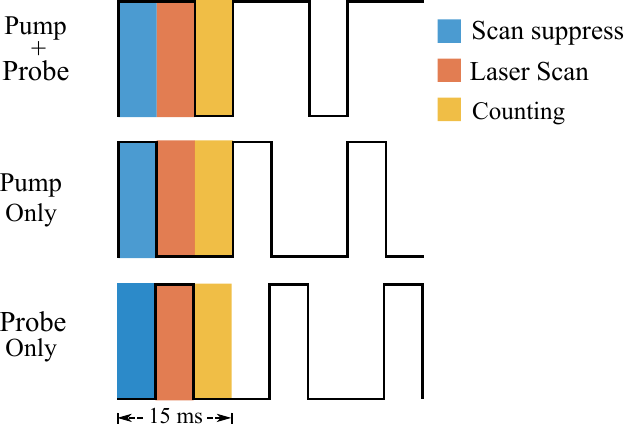}
    \caption{Timing sequence for the spectral hole burning spectroscopy. The duration AOM is turned on/off is shown by the black lines for three experiments with both pump-probe, pump-only, and probe-only light.}
    \label{fig:shb}
\end{figure}

For performing spectral hole burning experiments, we use sequences with three different variations, allowing us to use only a single laser. All sequences are divided into three 5 ms time windows: pump, probe, and count. During the pump window, the triangle wave is disconnected from the piezo and the laser is at a fixed wavelength, while for the probe and count windows, the piezo scans as normal. To be consistent, the piezo control always features these three windows; however, we switch the AOM to be on for only the pump, only the probe, or both, as shown in Fig. \ref{fig:shb}. In fact, there is a lag of 200 $\mu$s after the laser turns off before counting so that the excitation is not directly detected.

\section{Results and Discussion}

 \begin{figure}
    \centering
    \includegraphics[width=\columnwidth]{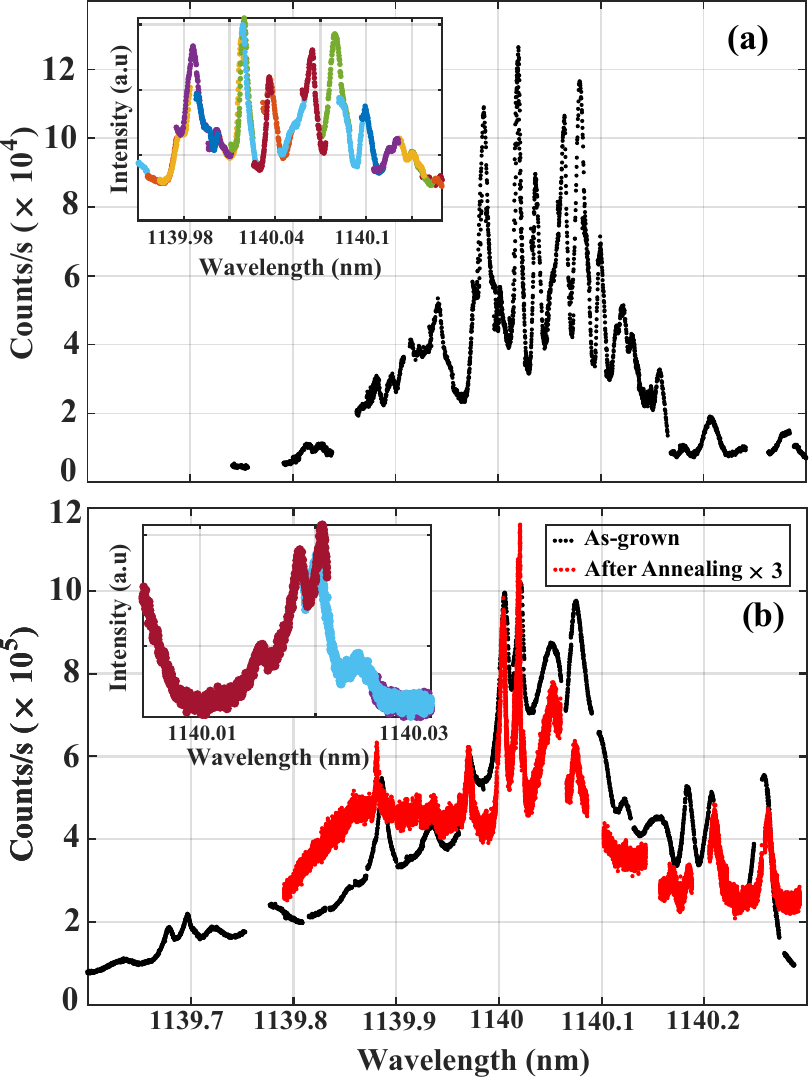}
    \caption{(a)Absorption spectrum of thulium trapped in neon. The inset shows a zoomed in view of the spectrum near 1140 nm, where each color represents one of the individual scans that are stitched together. (b) Absorption spectrum of thulium atoms trapped in an argon sample as grown (black) and in another sample after annealing (red). The inset shows a zoomed in view of the annealed sample showing features near 1140.02 nm. The data from the annealed sample is scaled by a factor of 3 for better comparison. }
    \label{fig:absorption}
\end{figure}

Using absorption spectroscopy and detecting all emitted infrared light, we obtain the spectrum shown in Fig. \ref{fig:absorption}. Because the mode-hop-free tuning range of the cateye laser is small, each of the color traces in the inset represents a separate 0.01 to 0.02 nm scan of the laser. Since large scans must be pieced together from many small scans, conditions vary slightly between acquisitions. Such conditions include laser power and precise alignment but may include other environmental factors. In order to present a more consistent spectrum, we have manually adjusted the spectra to overlap one another, which requires only very small corrections (by scaling up to 15\%). In both argon and neon, multiple narrow peaks are resolved. The presence of multiple peaks can be due to a multiplicity of trapping sites or a splitting of the ${}^2\mathrm{F}_{7/2}$ and ${}^2\mathrm{F}_{5/2}$ states into sub-levels.  Such sub-levels could result from the crystal field environment breaking rotation symmetry (crystal field splitting), from vibrational states of the trapped atom with its host ``cage'', or from hyperfine interactions with the nucleus. For vibrational states, if we estimate the potential energy landscape of a trapped thulium atom with the Ar-Ar Lennard-Jones potential\cite{dobbs_theory_1957}, we obtain an estimated vibrational frequency of 275 GHz or 1.2 nm, which is far too large to explain our results. Multiple peaks may also be due to the presence of other species - for example, ionized $\textrm{Tm}^+$ or $\textrm{Tm}^{2+}$, thulium dimers, clusters, or other compounds. However, $\textrm{Tm}^+$ has a different ground state fine structure due to coupling of the 6s electron with the f shell, which splits $J$ levels so that the only nearby line would be a dipole-forbidden $J = 4 \to J = 2$ transition at 1140.3 nm.
We find good evidence for both trapping site multiplicity and crystal field effects. In argon, the spectrum changes substantially as the temperature is raised and subsequently lowered (annealing) (see Fig. \ref{fig:absorption}), favoring the existence of multiple trapping sites that reconfigure during the anneal. Some variation of the spectrum with growth temperature is also observed, further supporting a model of multiple sites\cite{Dargyte2021}. Although Fig. \ref{fig:absorption} compares data between different samples, samples grown under the same conditions have reproducible spectra. Our emission spectrum (described below) reveals that excitation at one wavelength can yield emission at several others, suggesting the presence of sub-levels within the ${}^2\mathrm{F}_{7/2}$ and ${}^2\mathrm{F}_{5/2}$ states. However, considering each of the levels should be Kramers doublets, a maximum of 12 lines could be obtained - but more than these are apparent, possibly because each trapping site has its own set of crystal field levels. Unfortunately, due to the dense packing of so many lines, we could not identify specific sub-levels or measure their energy spacing. 

\begin{figure}
    \centering
    \includegraphics[width=\columnwidth]{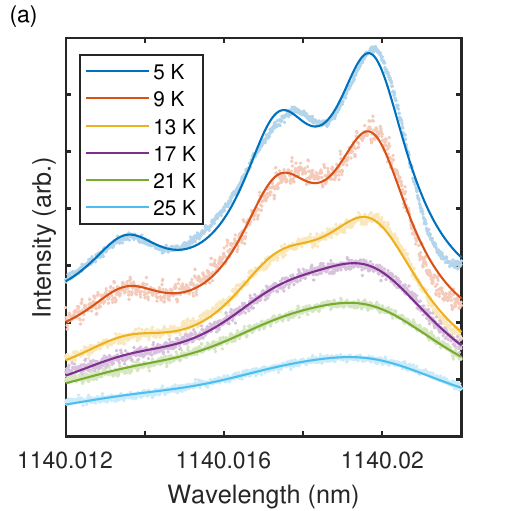}
    \includegraphics[width=\columnwidth]{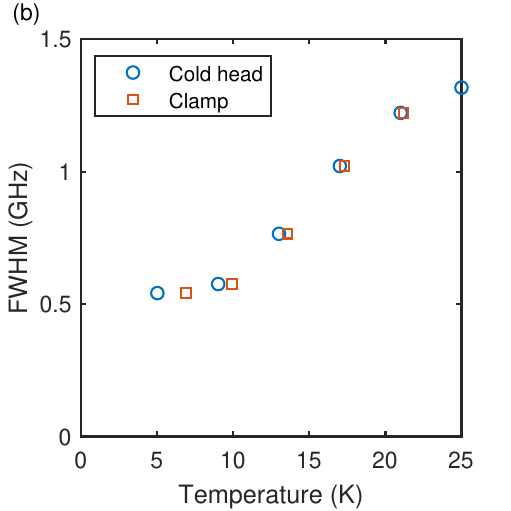}
    \caption{(a) Detailed spectrum in argon near 1140.02 nm taken at various temperatures (cold head temperature indicated). This sample was annealed to 25 K prior to the experiment, so the temperature dependence shown is reproducible. Solid lines are fits to a Lorentzian model. In this dataset only, a periodic noise source affected the detector, so a filter was applied to remove it. (b) Linewidth (FWHM) from the Lorentzian model as a function of temperature (both cold plate temperature and clamp temperature are shown). At the lowest temperatures, the thermal conductivity between components is limited and the sample temperature can't be precisely known.}
    \label{fig:temperature_fig}
\end{figure} 

The inset to Fig. \ref{fig:absorption}b shows the region near 1140.02 nm in detail. At this scale, additional structure is seen, with four identifiable peaks spaced by 0.5-1 GHz (0.002-0.004 nm). Such separations are comparable to the 1.5 GHz ground state hyperfine splitting in vacuum \cite{Giglberger1967} and may be provisionally assigned to a hyperfine origin. In a simple picture, both the ground and excited state Kramers doublets are split into two levels by the coupling to the nuclear spin, and there are then four transitions between them in a roughly symmetric pattern.

To further investigate the sources of broadening, we performed spectroscopy near 1140.02 nm at different temperatures. The results are shown in Fig. \ref{fig:temperature_fig}. At each temperature, the spectrum is fit to a sum of three Lorentzians and a background (the fourth peak observed in the inset of Fig. \ref{fig:absorption}b is just outside the scan). Note that the fit is constrained to have the locations of each peak fixed at all temperatures, and equal linewidth for all three peaks at a given temperature. This linewidth is plotted as a function of temperature in Fig. \ref{fig:temperature_fig}b, which shows the linewidth decreasing as temperature is reduced and supports our hypothesis that the linewidth is homogeneous thermal broadening from phonons. At temperatures much below the Debye temperature (67 K and 93 K for Ne and Ar, respectively\cite{pollack_solid_1964}), the temperature dependence of the linewidth is expected to be $T^7$ \cite{Friedrich1984,Konz2003}. It is not clear why the narrowing appears to saturate at lower temperatures or why this high power law dependence is not observed. While this could be an unknown additional broadening source, it may also be related to thermometry, as the two thermometers on the cold head and clamp bottom begin to deviate strongly as the temperature is reduced below 15 K, so the sample temperature may be significantly above either reading under the coldest conditions.

\begin{figure}
    \centering
    \includegraphics[width=\columnwidth]{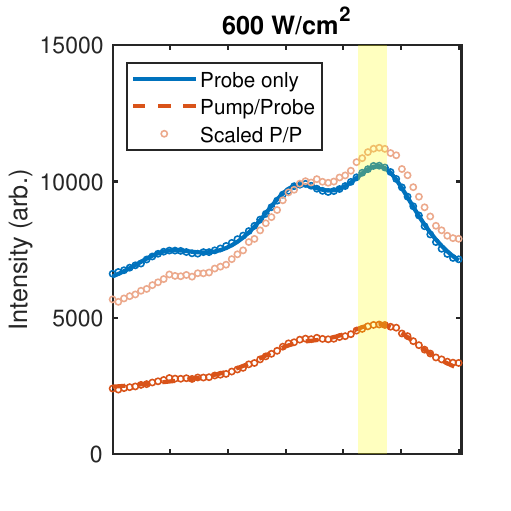}
    \includegraphics[width=\columnwidth]
    {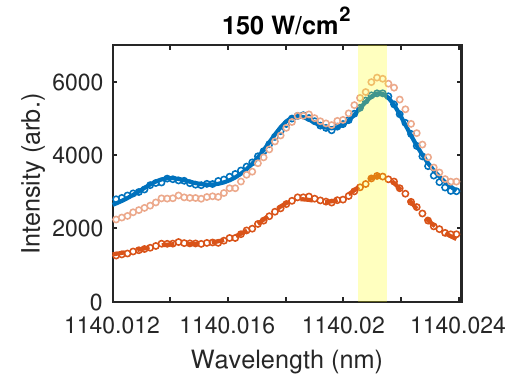}
   \caption{Spectral hole burning experiment in argon, performed with two different incident intensities. The yellow band indicates the wavelength of the pump. Note that the pump and probe are distinguished by the pump being applied first (each is applied for 5 ms), and are equal in power. We show three curves for each of two power levels: probe only (with background subtracted), pump/probe (the fluorescence from both together minus the pump-only fluorescence), and a rescaled version of pump/probe to compare with probe only. Solid lines are fits to a Lorentzian model.}
   \label{fig:shb_sat}
\end{figure}

We also attempted to find evidence for inhomogeneous broadening using a spectral hole burning strategy. For this experiment, the laser was jumped to a ``pump'' wavelength, then subsequently to a probe wavelength, then the light was turned off and the fluorescence recorded. These data are shown in Fig. \ref{fig:shb_sat}. We focus on the suppression caused by the pump on the probe fluorescence (and vice versa, since they are the same intensity), so we show the fluorescence of the probe alone (minus background), compared against the pump and probe together minus the pump alone (labeled pump/probe). For ease of comparison, we also show a scaled-up version of the pump/probe data. In the case of dominant inhomogeneous broadening, we expect the saturation to only affect those sites with nearby transition frequencies, leaving a suppressed signal, or ``hole,'' in the spectrum at the pump wavelength. However, no such effect is observed, and all lines are suppressed greatly over their entire width. This is observed for two different excitation powers. In fact, the line being pumped is actually suppressed less than neighboring lines, suggesting the dynamics are beyond a two-level model. This is also consistent with observation of significant saturation effects (four times the power gives only twice the fluorescence even without the pump) but rather small power broadening (linewidth increases in the fit by about 15\% as the power quadruples). Considering these facts, we conclude that these particular lines are split due to sub-levels (likely hyperfine) rather than multiplicity of trapping sites, and that the inhomogeneous broadening is too small to detect with present methods.

\begin{figure}
    \centering
    \includegraphics[width=\columnwidth]{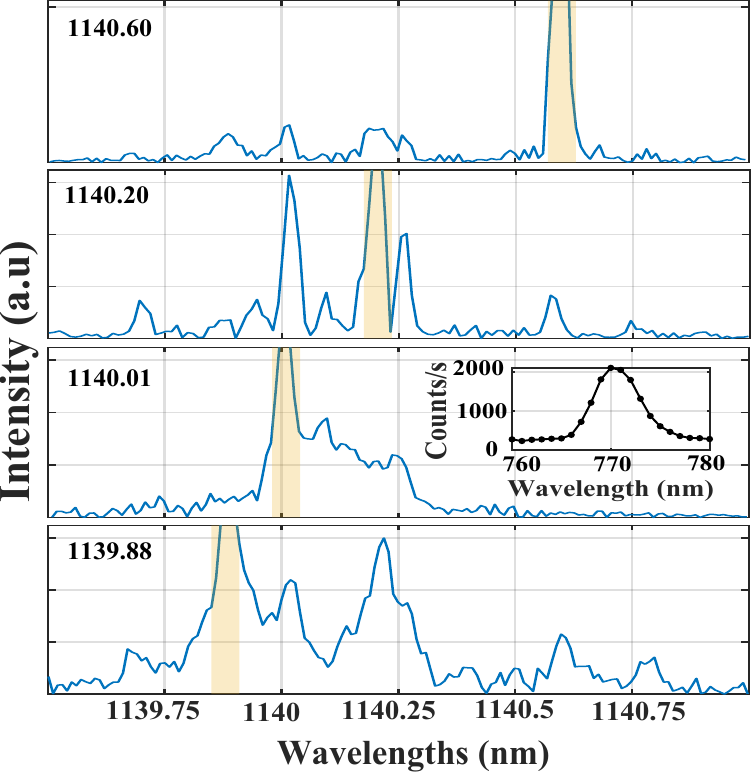}
    \caption{Emission features in Tm:Argon obtained by excitation at wavelengths given in each panel (shaded orange). Inset shows the emission observed near 770 nm for the same excitation as the main panel.}
    \label{fig:emis_spectra}
\end{figure}

Finally, we performed Fourier-transform spectroscopy on the fluorescence light. Fig. \ref{fig:emis_spectra} shows the emission spectra for four different excitation wavelengths. As can be seen, excitation at one wavelength yields emission at many others, supporting the existence of sub-levels from crystal field splitting. Both red- and blue-shifted lines are observed because the energy difference between these sub-levels is within the thermal spread. Common patterns are visible as well, as emission at 1140.6 nm occurs more strongly for excitation at 1139.88 nm or 1140.20 nm, and these wavelengths are also detected in the emission pattern when exciting at 1140.6 nm. If the only source of multiple lines were a multiplicity of trapping sites, then frequency-shifted emission would not be expected. Even if the neighboring atoms could exchange excitation by some inelastic mechanism, we would expect such a mechanism to predominantly occur for atoms with similar energies, yet for 1140.6 nm emission, one obtains stronger emission by exciting at 1139.88 nm than at 1140.01 nm. This pattern makes perfect sense in a combined trapping-site/sub-level model if there are trapping sites with sub-levels responsible for both 1140.6 nm and 1139.88 nm emission, while 1140.01 nm emission involves another site.

Although the sample is excited by infrared light, there is emission at visible wavelengths as well, and the sample can be seen by eye. This is likely due to atoms in the long-lived ${}^2\mathrm{F}_{5/2}$ state being further excited and decaying by an electric dipole allowed transition in the visible range. The majority of these emission wavelengths match known transitions in the 570-597 nm range, but an unknown emission line near 770 nm is present when the sample is excited with 1140.0 nm light. This could possibly be coming from a transition out of the $4f^{12}({}^3\textrm{H}_6)5d_{3/2}6s^2$ configuration with $J = 9/2$ into the ground ${}^2\mathrm{F}_{7/2}$ state, which would have a vacuum wavelength of $762\textrm{ nm}$.

\section{Conclusion}
Thulium atoms were embedded in crystals of argon and neon and we obtained laser-induced fluorescence spectra by excitation with infrared light. By exciting the magnetic dipole transition, it was observed in both the absorption and emission spectra that the transition was split into multiple components.  These splittings likely come from the crystal field of the argon or neon host, which breaks the rotational symmetry at the trapping site. The splitting of the $^2$F$_{7/2}$ and $^2$F$_{5/2}$ levels will depend on the detail of the symmetry breaking at the trapping site, which is not currently known. However, by a simple counting argument, we believe at least some lines originate from different trapping sites. Furthermore, we believe most of the linewidth is accounted for by homogeneous broadening, and there is no evidence from pump-probe experiments of any significant inhomogeneous broadening. This means that it should be possible to obtain still narrower lines by lowering the sample temperature. This, together with more sophisticated pump-probe experiments, should enable a complete characterization of the sub-levels of matrix-isolated thulium, paving the way for sensing and quantum information applications requiring a high density of identical emitters.

\bibliography{bibliographyFile.bib}

\begin{thebibliography}{31}%
\makeatletter
\providecommand \@ifxundefined [1]{%
 \@ifx{#1\undefined}
}%
\providecommand \@ifnum [1]{%
 \ifnum #1\expandafter \@firstoftwo
 \else \expandafter \@secondoftwo
 \fi
}%
\providecommand \@ifx [1]{%
 \ifx #1\expandafter \@firstoftwo
 \else \expandafter \@secondoftwo
 \fi
}%
\providecommand \natexlab [1]{#1}%
\providecommand \enquote  [1]{``#1''}%
\providecommand \bibnamefont  [1]{#1}%
\providecommand \bibfnamefont [1]{#1}%
\providecommand \citenamefont [1]{#1}%
\providecommand \href@noop [0]{\@secondoftwo}%
\providecommand \href [0]{\begingroup \@sanitize@url \@href}%
\providecommand \@href[1]{\@@startlink{#1}\@@href}%
\providecommand \@@href[1]{\endgroup#1\@@endlink}%
\providecommand \@sanitize@url [0]{\catcode `\\12\catcode `\$12\catcode
  `\&12\catcode `\#12\catcode `\^12\catcode `\_12\catcode `\%12\relax}%
\providecommand \@@startlink[1]{}%
\providecommand \@@endlink[0]{}%
\providecommand \url  [0]{\begingroup\@sanitize@url \@url }%
\providecommand \@url [1]{\endgroup\@href {#1}{\urlprefix }}%
\providecommand \urlprefix  [0]{URL }%
\providecommand \Eprint [0]{\href }%
\providecommand \doibase [0]{http://dx.doi.org/}%
\providecommand \selectlanguage [0]{\@gobble}%
\providecommand \bibinfo  [0]{\@secondoftwo}%
\providecommand \bibfield  [0]{\@secondoftwo}%
\providecommand \translation [1]{[#1]}%
\providecommand \BibitemOpen [0]{}%
\providecommand \bibitemStop [0]{}%
\providecommand \bibitemNoStop [0]{.\EOS\space}%
\providecommand \EOS [0]{\spacefactor3000\relax}%
\providecommand \BibitemShut  [1]{\csname bibitem#1\endcsname}%
\let\auto@bib@innerbib\@empty
\bibitem [{\citenamefont {Kitching}\ \emph {et~al.}(2011)\citenamefont
  {Kitching}, \citenamefont {Knappe},\ and\ \citenamefont
  {Donley}}]{kitching2011atomic}%
  \BibitemOpen
  \bibfield  {author} {\bibinfo {author} {\bibfnamefont {J.}~\bibnamefont
  {Kitching}}, \bibinfo {author} {\bibfnamefont {S.}~\bibnamefont {Knappe}}, \
  and\ \bibinfo {author} {\bibfnamefont {E.~A.}\ \bibnamefont {Donley}},\
  }\href@noop {} {\bibfield  {journal} {\bibinfo  {journal} {IEEE Sensors
  Journal}\ }\textbf {\bibinfo {volume} {11}},\ \bibinfo {pages} {1749}
  (\bibinfo {year} {2011})}\BibitemShut {NoStop}%
\bibitem [{\citenamefont {Ushijima}\ \emph {et~al.}(2015)\citenamefont
  {Ushijima}, \citenamefont {Takamoto}, \citenamefont {Das}, \citenamefont
  {Ohkubo},\ and\ \citenamefont {Katori}}]{ushijima2015cryogenic}%
  \BibitemOpen
  \bibfield  {author} {\bibinfo {author} {\bibfnamefont {I.}~\bibnamefont
  {Ushijima}}, \bibinfo {author} {\bibfnamefont {M.}~\bibnamefont {Takamoto}},
  \bibinfo {author} {\bibfnamefont {M.}~\bibnamefont {Das}}, \bibinfo {author}
  {\bibfnamefont {T.}~\bibnamefont {Ohkubo}}, \ and\ \bibinfo {author}
  {\bibfnamefont {H.}~\bibnamefont {Katori}},\ }\href
  {https://doi.org/10.1038/nphoton.2015.5} {\bibfield  {journal} {\bibinfo
  {journal} {Nature Photonics}\ }\textbf {\bibinfo {volume} {9}},\ \bibinfo
  {pages} {185} (\bibinfo {year} {2015})}\BibitemShut {NoStop}%
\bibitem [{\citenamefont {Camparo}(2007)}]{Camparo2007}%
  \BibitemOpen
  \bibfield  {author} {\bibinfo {author} {\bibfnamefont {J.}~\bibnamefont
  {Camparo}},\ }\href {\doibase 10.1063/1.2812121} {\bibfield  {journal}
  {\bibinfo  {journal} {Physics Today}\ }\textbf {\bibinfo {volume} {60}},\
  \bibinfo {pages} {33} (\bibinfo {year} {2007})},\ \Eprint
  {http://arxiv.org/abs/https://doi.org/10.1063/1.2812121}
  {https://doi.org/10.1063/1.2812121} \BibitemShut {NoStop}%
\bibitem [{\citenamefont {Biedermann}\ \emph {et~al.}(2017)\citenamefont
  {Biedermann}, \citenamefont {McGuinness}, \citenamefont {Rakholia},
  \citenamefont {Jau}, \citenamefont {Wheeler}, \citenamefont {Sterk},\ and\
  \citenamefont {Burns}}]{Biedermann2017}%
  \BibitemOpen
  \bibfield  {author} {\bibinfo {author} {\bibfnamefont {G.~W.}\ \bibnamefont
  {Biedermann}}, \bibinfo {author} {\bibfnamefont {H.~J.}\ \bibnamefont
  {McGuinness}}, \bibinfo {author} {\bibfnamefont {A.~V.}\ \bibnamefont
  {Rakholia}}, \bibinfo {author} {\bibfnamefont {Y.-Y.}\ \bibnamefont {Jau}},
  \bibinfo {author} {\bibfnamefont {D.~R.}\ \bibnamefont {Wheeler}}, \bibinfo
  {author} {\bibfnamefont {J.~D.}\ \bibnamefont {Sterk}}, \ and\ \bibinfo
  {author} {\bibfnamefont {G.~R.}\ \bibnamefont {Burns}},\ }\href {\doibase
  10.1103/PhysRevLett.118.163601} {\bibfield  {journal} {\bibinfo  {journal}
  {Phys. Rev. Lett.}\ }\textbf {\bibinfo {volume} {118}},\ \bibinfo {pages}
  {163601} (\bibinfo {year} {2017})}\BibitemShut {NoStop}%
\bibitem [{\citenamefont {Charri\`ere}\ \emph {et~al.}(2012)\citenamefont
  {Charri\`ere}, \citenamefont {Cadoret}, \citenamefont {Zahzam}, \citenamefont
  {Bidel},\ and\ \citenamefont {Bresson}}]{Charriere2012}%
  \BibitemOpen
  \bibfield  {author} {\bibinfo {author} {\bibfnamefont {R.}~\bibnamefont
  {Charri\`ere}}, \bibinfo {author} {\bibfnamefont {M.}~\bibnamefont
  {Cadoret}}, \bibinfo {author} {\bibfnamefont {N.}~\bibnamefont {Zahzam}},
  \bibinfo {author} {\bibfnamefont {Y.}~\bibnamefont {Bidel}}, \ and\ \bibinfo
  {author} {\bibfnamefont {A.}~\bibnamefont {Bresson}},\ }\href {\doibase
  10.1103/PhysRevA.85.013639} {\bibfield  {journal} {\bibinfo  {journal} {Phys.
  Rev. A}\ }\textbf {\bibinfo {volume} {85}},\ \bibinfo {pages} {013639}
  (\bibinfo {year} {2012})}\BibitemShut {NoStop}%
\bibitem [{\citenamefont {McGuinness}\ \emph {et~al.}(2012)\citenamefont
  {McGuinness}, \citenamefont {Rakholia},\ and\ \citenamefont
  {Biedermann}}]{Mcguinness2012}%
  \BibitemOpen
  \bibfield  {author} {\bibinfo {author} {\bibfnamefont {H.~J.}\ \bibnamefont
  {McGuinness}}, \bibinfo {author} {\bibfnamefont {A.~V.}\ \bibnamefont
  {Rakholia}}, \ and\ \bibinfo {author} {\bibfnamefont {G.~W.}\ \bibnamefont
  {Biedermann}},\ }\href {\doibase 10.1063/1.3673845} {\bibfield  {journal}
  {\bibinfo  {journal} {Applied Physics Letters}\ }\textbf {\bibinfo {volume}
  {100}},\ \bibinfo {pages} {011106} (\bibinfo {year} {2012})},\ \Eprint
  {http://arxiv.org/abs/https://doi.org/10.1063/1.3673845}
  {https://doi.org/10.1063/1.3673845} \BibitemShut {NoStop}%
\bibitem [{\citenamefont {Wickenbrock}\ \emph {et~al.}(2013)\citenamefont
  {Wickenbrock}, \citenamefont {Tricot},\ and\ \citenamefont
  {Renzoni}}]{Wickenbrock2013}%
  \BibitemOpen
  \bibfield  {author} {\bibinfo {author} {\bibfnamefont {A.}~\bibnamefont
  {Wickenbrock}}, \bibinfo {author} {\bibfnamefont {F.}~\bibnamefont {Tricot}},
  \ and\ \bibinfo {author} {\bibfnamefont {F.}~\bibnamefont {Renzoni}},\ }\href
  {\doibase 10.1063/1.4848196} {\bibfield  {journal} {\bibinfo  {journal}
  {Applied Physics Letters}\ }\textbf {\bibinfo {volume} {103}},\ \bibinfo
  {pages} {243503} (\bibinfo {year} {2013})},\ \Eprint
  {http://arxiv.org/abs/https://doi.org/10.1063/1.4848196}
  {https://doi.org/10.1063/1.4848196} \BibitemShut {NoStop}%
\bibitem [{\citenamefont {Patton}\ \emph {et~al.}(2014)\citenamefont {Patton},
  \citenamefont {Zhivun}, \citenamefont {Hovde},\ and\ \citenamefont
  {Budker}}]{Patton2014}%
  \BibitemOpen
  \bibfield  {author} {\bibinfo {author} {\bibfnamefont {B.}~\bibnamefont
  {Patton}}, \bibinfo {author} {\bibfnamefont {E.}~\bibnamefont {Zhivun}},
  \bibinfo {author} {\bibfnamefont {D.~C.}\ \bibnamefont {Hovde}}, \ and\
  \bibinfo {author} {\bibfnamefont {D.}~\bibnamefont {Budker}},\ }\href
  {\doibase 10.1103/PhysRevLett.113.013001} {\bibfield  {journal} {\bibinfo
  {journal} {Phys. Rev. Lett.}\ }\textbf {\bibinfo {volume} {113}},\ \bibinfo
  {pages} {013001} (\bibinfo {year} {2014})}\BibitemShut {NoStop}%
\bibitem [{\citenamefont {Zotev}\ \emph {et~al.}(2007)\citenamefont {Zotev},
  \citenamefont {Matlashov}, \citenamefont {Volegov}, \citenamefont {Urbaitis},
  \citenamefont {Espy},\ and\ \citenamefont {Jr}}]{Zotev_2007}%
  \BibitemOpen
  \bibfield  {author} {\bibinfo {author} {\bibfnamefont {V.~S.}\ \bibnamefont
  {Zotev}}, \bibinfo {author} {\bibfnamefont {A.~N.}\ \bibnamefont
  {Matlashov}}, \bibinfo {author} {\bibfnamefont {P.~L.}\ \bibnamefont
  {Volegov}}, \bibinfo {author} {\bibfnamefont {A.~V.}\ \bibnamefont
  {Urbaitis}}, \bibinfo {author} {\bibfnamefont {M.~A.}\ \bibnamefont {Espy}},
  \ and\ \bibinfo {author} {\bibfnamefont {R.~H.~K.}\ \bibnamefont {Jr}},\
  }\href {\doibase 10.1088/0953-2048/20/11/s13} {\bibfield  {journal} {\bibinfo
   {journal} {Superconductor Science and Technology}\ }\textbf {\bibinfo
  {volume} {20}},\ \bibinfo {pages} {S367} (\bibinfo {year}
  {2007})}\BibitemShut {NoStop}%
\bibitem [{\citenamefont {Matlachov}\ \emph {et~al.}(2004)\citenamefont
  {Matlachov}, \citenamefont {Volegov}, \citenamefont {Espy}, \citenamefont
  {George},\ and\ \citenamefont {Kraus}}]{MATLACHOV20041}%
  \BibitemOpen
  \bibfield  {author} {\bibinfo {author} {\bibfnamefont {A.~N.}\ \bibnamefont
  {Matlachov}}, \bibinfo {author} {\bibfnamefont {P.~L.}\ \bibnamefont
  {Volegov}}, \bibinfo {author} {\bibfnamefont {M.~A.}\ \bibnamefont {Espy}},
  \bibinfo {author} {\bibfnamefont {J.~S.}\ \bibnamefont {George}}, \ and\
  \bibinfo {author} {\bibfnamefont {R.~H.}\ \bibnamefont {Kraus}},\ }\href
  {\doibase https://doi.org/10.1016/j.jmr.2004.05.015} {\bibfield  {journal}
  {\bibinfo  {journal} {Journal of Magnetic Resonance}\ }\textbf {\bibinfo
  {volume} {170}},\ \bibinfo {pages} {1} (\bibinfo {year} {2004})}\BibitemShut
  {NoStop}%
\bibitem [{\citenamefont {Gatteschi}\ and\ \citenamefont
  {Sessoli}(2003)}]{Gatteschi2003}%
  \BibitemOpen
  \bibfield  {author} {\bibinfo {author} {\bibfnamefont {D.}~\bibnamefont
  {Gatteschi}}\ and\ \bibinfo {author} {\bibfnamefont {R.}~\bibnamefont
  {Sessoli}},\ }\href {\doibase https://doi.org/10.1002/anie.200390099}
  {\bibfield  {journal} {\bibinfo  {journal} {Angewandte Chemie International
  Edition}\ }\textbf {\bibinfo {volume} {42}},\ \bibinfo {pages} {268}
  (\bibinfo {year} {2003})},\ \Eprint
  {http://arxiv.org/abs/https://onlinelibrary.wiley.com/doi/pdf/10.1002/anie.200390099}
  {https://onlinelibrary.wiley.com/doi/pdf/10.1002/anie.200390099} \BibitemShut
  {NoStop}%
\bibitem [{\citenamefont {Sawicki}\ \emph {et~al.}(2011)\citenamefont
  {Sawicki}, \citenamefont {Stefanowicz},\ and\ \citenamefont
  {Ney}}]{Sawicki2011}%
  \BibitemOpen
  \bibfield  {author} {\bibinfo {author} {\bibfnamefont {M.}~\bibnamefont
  {Sawicki}}, \bibinfo {author} {\bibfnamefont {W.}~\bibnamefont
  {Stefanowicz}}, \ and\ \bibinfo {author} {\bibfnamefont {A.}~\bibnamefont
  {Ney}},\ }\href {\doibase 10.1088/0268-1242/26/6/064006} {\bibfield
  {journal} {\bibinfo  {journal} {Semiconductor Science and Technology}\
  }\textbf {\bibinfo {volume} {26}},\ \bibinfo {pages} {064006} (\bibinfo
  {year} {2011})}\BibitemShut {NoStop}%
\bibitem [{\citenamefont {Barry}\ \emph {et~al.}(2020)\citenamefont {Barry},
  \citenamefont {Schloss}, \citenamefont {Bauch}, \citenamefont {Turner},
  \citenamefont {Hart}, \citenamefont {Pham},\ and\ \citenamefont
  {Walsworth}}]{barry2020}%
  \BibitemOpen
  \bibfield  {author} {\bibinfo {author} {\bibfnamefont {J.~F.}\ \bibnamefont
  {Barry}}, \bibinfo {author} {\bibfnamefont {J.~M.}\ \bibnamefont {Schloss}},
  \bibinfo {author} {\bibfnamefont {E.}~\bibnamefont {Bauch}}, \bibinfo
  {author} {\bibfnamefont {M.~J.}\ \bibnamefont {Turner}}, \bibinfo {author}
  {\bibfnamefont {C.~A.}\ \bibnamefont {Hart}}, \bibinfo {author}
  {\bibfnamefont {L.~M.}\ \bibnamefont {Pham}}, \ and\ \bibinfo {author}
  {\bibfnamefont {R.~L.}\ \bibnamefont {Walsworth}},\ }\href {\doibase
  10.1103/RevModPhys.92.015004} {\bibfield  {journal} {\bibinfo  {journal}
  {Rev. Mod. Phys.}\ }\textbf {\bibinfo {volume} {92}},\ \bibinfo {pages}
  {015004} (\bibinfo {year} {2020})}\BibitemShut {NoStop}%
\bibitem [{\citenamefont {Lesik}\ \emph {et~al.}(2019)\citenamefont {Lesik},
  \citenamefont {Plisson}, \citenamefont {Toraille}, \citenamefont {Renaud},
  \citenamefont {Occelli}, \citenamefont {Schmidt}, \citenamefont {Salord},
  \citenamefont {Delobbe}, \citenamefont {Debuisschert}, \citenamefont {Rondin}
  \emph {et~al.}}]{lesik2019magnetic}%
  \BibitemOpen
  \bibfield  {author} {\bibinfo {author} {\bibfnamefont {M.}~\bibnamefont
  {Lesik}}, \bibinfo {author} {\bibfnamefont {T.}~\bibnamefont {Plisson}},
  \bibinfo {author} {\bibfnamefont {L.}~\bibnamefont {Toraille}}, \bibinfo
  {author} {\bibfnamefont {J.}~\bibnamefont {Renaud}}, \bibinfo {author}
  {\bibfnamefont {F.}~\bibnamefont {Occelli}}, \bibinfo {author} {\bibfnamefont
  {M.}~\bibnamefont {Schmidt}}, \bibinfo {author} {\bibfnamefont
  {O.}~\bibnamefont {Salord}}, \bibinfo {author} {\bibfnamefont
  {A.}~\bibnamefont {Delobbe}}, \bibinfo {author} {\bibfnamefont
  {T.}~\bibnamefont {Debuisschert}}, \bibinfo {author} {\bibfnamefont
  {L.}~\bibnamefont {Rondin}},  \emph {et~al.},\ }\href@noop {} {\bibfield
  {journal} {\bibinfo  {journal} {Science}\ }\textbf {\bibinfo {volume}
  {366}},\ \bibinfo {pages} {1359} (\bibinfo {year} {2019})}\BibitemShut
  {NoStop}%
\bibitem [{\citenamefont {Childress}\ and\ \citenamefont
  {Hanson}(2013)}]{childress2013}%
  \BibitemOpen
  \bibfield  {author} {\bibinfo {author} {\bibfnamefont {L.}~\bibnamefont
  {Childress}}\ and\ \bibinfo {author} {\bibfnamefont {R.}~\bibnamefont
  {Hanson}},\ }\href {\doibase 10.1557/mrs.2013.20} {\bibfield  {journal}
  {\bibinfo  {journal} {MRS Bulletin}\ }\textbf {\bibinfo {volume} {38}},\
  \bibinfo {pages} {134–138} (\bibinfo {year} {2013})}\BibitemShut {NoStop}%
\bibitem [{\citenamefont {Acosta}\ \emph {et~al.}(2009)\citenamefont {Acosta},
  \citenamefont {Bauch}, \citenamefont {Ledbetter}, \citenamefont {Santori},
  \citenamefont {Fu}, \citenamefont {Barclay}, \citenamefont {Beausoleil},
  \citenamefont {Linget}, \citenamefont {Roch}, \citenamefont {Treussart},
  \citenamefont {Chemerisov}, \citenamefont {Gawlik},\ and\ \citenamefont
  {Budker}}]{Acosta2009}%
  \BibitemOpen
  \bibfield  {author} {\bibinfo {author} {\bibfnamefont {V.~M.}\ \bibnamefont
  {Acosta}}, \bibinfo {author} {\bibfnamefont {E.}~\bibnamefont {Bauch}},
  \bibinfo {author} {\bibfnamefont {M.~P.}\ \bibnamefont {Ledbetter}}, \bibinfo
  {author} {\bibfnamefont {C.}~\bibnamefont {Santori}}, \bibinfo {author}
  {\bibfnamefont {K.-M.~C.}\ \bibnamefont {Fu}}, \bibinfo {author}
  {\bibfnamefont {P.~E.}\ \bibnamefont {Barclay}}, \bibinfo {author}
  {\bibfnamefont {R.~G.}\ \bibnamefont {Beausoleil}}, \bibinfo {author}
  {\bibfnamefont {H.}~\bibnamefont {Linget}}, \bibinfo {author} {\bibfnamefont
  {J.~F.}\ \bibnamefont {Roch}}, \bibinfo {author} {\bibfnamefont
  {F.}~\bibnamefont {Treussart}}, \bibinfo {author} {\bibfnamefont
  {S.}~\bibnamefont {Chemerisov}}, \bibinfo {author} {\bibfnamefont
  {W.}~\bibnamefont {Gawlik}}, \ and\ \bibinfo {author} {\bibfnamefont
  {D.}~\bibnamefont {Budker}},\ }\href {\doibase 10.1103/PhysRevB.80.115202}
  {\bibfield  {journal} {\bibinfo  {journal} {Phys. Rev. B}\ }\textbf {\bibinfo
  {volume} {80}},\ \bibinfo {pages} {115202} (\bibinfo {year}
  {2009})}\BibitemShut {NoStop}%
\bibitem [{\citenamefont {Taylor}\ \emph {et~al.}(2008)\citenamefont {Taylor},
  \citenamefont {Cappellaro}, \citenamefont {Childress}, \citenamefont {Jiang},
  \citenamefont {Budker}, \citenamefont {Hemmer}, \citenamefont {Yacoby},
  \citenamefont {Walsworth},\ and\ \citenamefont {Lukin}}]{Taylor2008}%
  \BibitemOpen
  \bibfield  {author} {\bibinfo {author} {\bibfnamefont {J.~M.}\ \bibnamefont
  {Taylor}}, \bibinfo {author} {\bibfnamefont {P.}~\bibnamefont {Cappellaro}},
  \bibinfo {author} {\bibfnamefont {L.}~\bibnamefont {Childress}}, \bibinfo
  {author} {\bibfnamefont {L.}~\bibnamefont {Jiang}}, \bibinfo {author}
  {\bibfnamefont {D.}~\bibnamefont {Budker}}, \bibinfo {author} {\bibfnamefont
  {P.~R.}\ \bibnamefont {Hemmer}}, \bibinfo {author} {\bibfnamefont
  {A.}~\bibnamefont {Yacoby}}, \bibinfo {author} {\bibfnamefont
  {R.}~\bibnamefont {Walsworth}}, \ and\ \bibinfo {author} {\bibfnamefont
  {M.~D.}\ \bibnamefont {Lukin}},\ }\href {\doibase 10.1038/nphys1075}
  {\bibfield  {journal} {\bibinfo  {journal} {Nature Physics}\ }\textbf
  {\bibinfo {volume} {4}},\ \bibinfo {pages} {810} (\bibinfo {year}
  {2008})}\BibitemShut {NoStop}%
\bibitem [{\citenamefont {Rogers}\ \emph {et~al.}(2014)\citenamefont {Rogers},
  \citenamefont {Jahnke}, \citenamefont {Teraji}, \citenamefont {Marseglia},
  \citenamefont {Müller}, \citenamefont {Naydenov}, \citenamefont
  {Schauffert}, \citenamefont {Kranz}, \citenamefont {Isoya}, \citenamefont
  {McGuinness},\ and\ \citenamefont {Jelezko}}]{rogers_multiple_2014}%
  \BibitemOpen
  \bibfield  {author} {\bibinfo {author} {\bibfnamefont {L.}~\bibnamefont
  {Rogers}}, \bibinfo {author} {\bibfnamefont {K.}~\bibnamefont {Jahnke}},
  \bibinfo {author} {\bibfnamefont {T.}~\bibnamefont {Teraji}}, \bibinfo
  {author} {\bibfnamefont {L.}~\bibnamefont {Marseglia}}, \bibinfo {author}
  {\bibfnamefont {C.}~\bibnamefont {Müller}}, \bibinfo {author} {\bibfnamefont
  {B.}~\bibnamefont {Naydenov}}, \bibinfo {author} {\bibfnamefont
  {H.}~\bibnamefont {Schauffert}}, \bibinfo {author} {\bibfnamefont
  {C.}~\bibnamefont {Kranz}}, \bibinfo {author} {\bibfnamefont
  {J.}~\bibnamefont {Isoya}}, \bibinfo {author} {\bibfnamefont
  {L.}~\bibnamefont {McGuinness}}, \ and\ \bibinfo {author} {\bibfnamefont
  {F.}~\bibnamefont {Jelezko}},\ }\href {\doibase 10.1038/ncomms5739}
  {\bibfield  {journal} {\bibinfo  {journal} {Nature Communications}\ }\textbf
  {\bibinfo {volume} {5}},\ \bibinfo {pages} {4739} (\bibinfo {year}
  {2014})}\BibitemShut {NoStop}%
\bibitem [{\citenamefont {van Dam}\ \emph {et~al.}(2019)\citenamefont {van
  Dam}, \citenamefont {Walsh}, \citenamefont {Degen}, \citenamefont {Bersin},
  \citenamefont {Mouradian}, \citenamefont {Galiullin}, \citenamefont {Ruf},
  \citenamefont {IJspeert}, \citenamefont {Taminiau}, \citenamefont {Hanson},\
  and\ \citenamefont {Englund}}]{Dam2019}%
  \BibitemOpen
  \bibfield  {author} {\bibinfo {author} {\bibfnamefont {S.~B.}\ \bibnamefont
  {van Dam}}, \bibinfo {author} {\bibfnamefont {M.}~\bibnamefont {Walsh}},
  \bibinfo {author} {\bibfnamefont {M.~J.}\ \bibnamefont {Degen}}, \bibinfo
  {author} {\bibfnamefont {E.}~\bibnamefont {Bersin}}, \bibinfo {author}
  {\bibfnamefont {S.~L.}\ \bibnamefont {Mouradian}}, \bibinfo {author}
  {\bibfnamefont {A.}~\bibnamefont {Galiullin}}, \bibinfo {author}
  {\bibfnamefont {M.}~\bibnamefont {Ruf}}, \bibinfo {author} {\bibfnamefont
  {M.}~\bibnamefont {IJspeert}}, \bibinfo {author} {\bibfnamefont {T.~H.}\
  \bibnamefont {Taminiau}}, \bibinfo {author} {\bibfnamefont {R.}~\bibnamefont
  {Hanson}}, \ and\ \bibinfo {author} {\bibfnamefont {D.~R.}\ \bibnamefont
  {Englund}},\ }\href {\doibase 10.1103/PhysRevB.99.161203} {\bibfield
  {journal} {\bibinfo  {journal} {Phys. Rev. B}\ }\textbf {\bibinfo {volume}
  {99}},\ \bibinfo {pages} {161203} (\bibinfo {year} {2019})}\BibitemShut
  {NoStop}%
\bibitem [{\citenamefont {Kanagin}\ \emph {et~al.}(2013)\citenamefont
  {Kanagin}, \citenamefont {Regmi}, \citenamefont {Pathak},\ and\ \citenamefont
  {Weinstein}}]{kanagin2013optical}%
  \BibitemOpen
  \bibfield  {author} {\bibinfo {author} {\bibfnamefont {A.~N.}\ \bibnamefont
  {Kanagin}}, \bibinfo {author} {\bibfnamefont {S.~K.}\ \bibnamefont {Regmi}},
  \bibinfo {author} {\bibfnamefont {P.}~\bibnamefont {Pathak}}, \ and\ \bibinfo
  {author} {\bibfnamefont {J.~D.}\ \bibnamefont {Weinstein}},\ }\href@noop {}
  {\bibfield  {journal} {\bibinfo  {journal} {Physical Review A}\ }\textbf
  {\bibinfo {volume} {88}},\ \bibinfo {pages} {063404} (\bibinfo {year}
  {2013})}\BibitemShut {NoStop}%
\bibitem [{\citenamefont {Gaire}\ \emph {et~al.}(2019)\citenamefont {Gaire},
  \citenamefont {Raman},\ and\ \citenamefont {Parker}}]{gaire2019}%
  \BibitemOpen
  \bibfield  {author} {\bibinfo {author} {\bibfnamefont {V.}~\bibnamefont
  {Gaire}}, \bibinfo {author} {\bibfnamefont {C.~S.}\ \bibnamefont {Raman}}, \
  and\ \bibinfo {author} {\bibfnamefont {C.~V.}\ \bibnamefont {Parker}},\
  }\href {\doibase 10.1103/PhysRevA.99.022505} {\bibfield  {journal} {\bibinfo
  {journal} {Phys. Rev. A}\ }\textbf {\bibinfo {volume} {99}},\ \bibinfo
  {pages} {022505} (\bibinfo {year} {2019})}\BibitemShut {NoStop}%
\bibitem [{\citenamefont {Ishikawa}\ \emph {et~al.}(1997)\citenamefont
  {Ishikawa}, \citenamefont {Hatakeyama}, \citenamefont {Gosyono-o},
  \citenamefont {Wada}, \citenamefont {Takahashi},\ and\ \citenamefont
  {Yabuzaki}}]{ishikawa1997}%
  \BibitemOpen
  \bibfield  {author} {\bibinfo {author} {\bibfnamefont {K.}~\bibnamefont
  {Ishikawa}}, \bibinfo {author} {\bibfnamefont {A.}~\bibnamefont
  {Hatakeyama}}, \bibinfo {author} {\bibfnamefont {K.}~\bibnamefont
  {Gosyono-o}}, \bibinfo {author} {\bibfnamefont {S.}~\bibnamefont {Wada}},
  \bibinfo {author} {\bibfnamefont {Y.}~\bibnamefont {Takahashi}}, \ and\
  \bibinfo {author} {\bibfnamefont {T.}~\bibnamefont {Yabuzaki}},\ }\href
  {\doibase 10.1103/PhysRevB.56.780} {\bibfield  {journal} {\bibinfo  {journal}
  {Phys. Rev. B}\ }\textbf {\bibinfo {volume} {56}},\ \bibinfo {pages} {780}
  (\bibinfo {year} {1997})}\BibitemShut {NoStop}%
\bibitem [{\citenamefont {Singh}\ \emph {et~al.}(2020)\citenamefont {Singh},
  \citenamefont {Prakash}, \citenamefont {Wolfowicz}, \citenamefont {Wen},
  \citenamefont {Huang}, \citenamefont {Rajh}, \citenamefont {Awschalom},
  \citenamefont {Zhong},\ and\ \citenamefont {Guha}}]{Singh2020}%
  \BibitemOpen
  \bibfield  {author} {\bibinfo {author} {\bibfnamefont {M.~K.}\ \bibnamefont
  {Singh}}, \bibinfo {author} {\bibfnamefont {A.}~\bibnamefont {Prakash}},
  \bibinfo {author} {\bibfnamefont {G.}~\bibnamefont {Wolfowicz}}, \bibinfo
  {author} {\bibfnamefont {J.}~\bibnamefont {Wen}}, \bibinfo {author}
  {\bibfnamefont {Y.}~\bibnamefont {Huang}}, \bibinfo {author} {\bibfnamefont
  {T.}~\bibnamefont {Rajh}}, \bibinfo {author} {\bibfnamefont {D.~D.}\
  \bibnamefont {Awschalom}}, \bibinfo {author} {\bibfnamefont {T.}~\bibnamefont
  {Zhong}}, \ and\ \bibinfo {author} {\bibfnamefont {S.}~\bibnamefont {Guha}},\
  }\href {\doibase 10.1063/1.5142611} {\bibfield  {journal} {\bibinfo
  {journal} {APL Materials}\ }\textbf {\bibinfo {volume} {8}},\ \bibinfo
  {pages} {031111} (\bibinfo {year} {2020})},\ \Eprint
  {http://arxiv.org/abs/https://doi.org/10.1063/1.5142611}
  {https://doi.org/10.1063/1.5142611} \BibitemShut {NoStop}%
\bibitem [{\citenamefont {B\"ottger}\ \emph {et~al.}(2016)\citenamefont
  {B\"ottger}, \citenamefont {Thiel}, \citenamefont {Cone}, \citenamefont
  {Sun},\ and\ \citenamefont {Faraon}}]{Bottger2016}%
  \BibitemOpen
  \bibfield  {author} {\bibinfo {author} {\bibfnamefont {T.}~\bibnamefont
  {B\"ottger}}, \bibinfo {author} {\bibfnamefont {C.~W.}\ \bibnamefont
  {Thiel}}, \bibinfo {author} {\bibfnamefont {R.~L.}\ \bibnamefont {Cone}},
  \bibinfo {author} {\bibfnamefont {Y.}~\bibnamefont {Sun}}, \ and\ \bibinfo
  {author} {\bibfnamefont {A.}~\bibnamefont {Faraon}},\ }\href {\doibase
  10.1103/PhysRevB.94.045134} {\bibfield  {journal} {\bibinfo  {journal} {Phys.
  Rev. B}\ }\textbf {\bibinfo {volume} {94}},\ \bibinfo {pages} {045134}
  (\bibinfo {year} {2016})}\BibitemShut {NoStop}%
\bibitem [{\citenamefont {Gaire}\ and\ \citenamefont
  {Parker}(2022)}]{gaire2022_ftir}%
  \BibitemOpen
  \bibfield  {author} {\bibinfo {author} {\bibfnamefont {V.}~\bibnamefont
  {Gaire}}\ and\ \bibinfo {author} {\bibfnamefont {C.~V.}\ \bibnamefont
  {Parker}},\ }\href {\doibase 10.1364/JOSAA.458357} {\enquote {\bibinfo
  {title} {A self-calibrated fourier transform spectrometer for laser-induced
  fluorescence spectroscopy with single-photon avalanche diode detection},}\ }
  (\bibinfo {year} {2022}),\ \bibinfo {note} {(in press)}\BibitemShut {NoStop}%
\bibitem [{\citenamefont {Dobbs}\ and\ \citenamefont
  {Jones}(1957)}]{dobbs_theory_1957}%
  \BibitemOpen
  \bibfield  {author} {\bibinfo {author} {\bibfnamefont {E.~R.}\ \bibnamefont
  {Dobbs}}\ and\ \bibinfo {author} {\bibfnamefont {G.~O.}\ \bibnamefont
  {Jones}},\ }\href {\doibase 10.1088/0034-4885/20/1/309} {\bibfield  {journal}
  {\bibinfo  {journal} {Reports on Progress in Physics}\ }\textbf {\bibinfo
  {volume} {20}},\ \bibinfo {pages} {516} (\bibinfo {year} {1957})}\BibitemShut
  {NoStop}%
\bibitem [{\citenamefont {Dargyte}\ \emph {et~al.}(2021)\citenamefont
  {Dargyte}, \citenamefont {Lancaster},\ and\ \citenamefont
  {Weinstein}}]{Dargyte2021}%
  \BibitemOpen
  \bibfield  {author} {\bibinfo {author} {\bibfnamefont {U.}~\bibnamefont
  {Dargyte}}, \bibinfo {author} {\bibfnamefont {D.~M.}\ \bibnamefont
  {Lancaster}}, \ and\ \bibinfo {author} {\bibfnamefont {J.~D.}\ \bibnamefont
  {Weinstein}},\ }\href {\doibase 10.1103/PhysRevA.104.032611} {\bibfield
  {journal} {\bibinfo  {journal} {Phys. Rev. A}\ }\textbf {\bibinfo {volume}
  {104}},\ \bibinfo {pages} {032611} (\bibinfo {year} {2021})}\BibitemShut
  {NoStop}%
\bibitem [{\citenamefont {Giglberger}\ and\ \citenamefont
  {Penselin}(1967)}]{Giglberger1967}%
  \BibitemOpen
  \bibfield  {author} {\bibinfo {author} {\bibfnamefont {D.}~\bibnamefont
  {Giglberger}}\ and\ \bibinfo {author} {\bibfnamefont {S.}~\bibnamefont
  {Penselin}},\ }\href {\doibase 10.1007/BF01326434} {\bibfield  {journal}
  {\bibinfo  {journal} {Zeitschrift f{\"u}r Physik}\ }\textbf {\bibinfo
  {volume} {199}},\ \bibinfo {pages} {244} (\bibinfo {year}
  {1967})}\BibitemShut {NoStop}%
\bibitem [{\citenamefont {Pollack}(1964)}]{pollack_solid_1964}%
  \BibitemOpen
  \bibfield  {author} {\bibinfo {author} {\bibfnamefont {G.~L.}\ \bibnamefont
  {Pollack}},\ }\href {\doibase 10.1103/RevModPhys.36.748} {\bibfield
  {journal} {\bibinfo  {journal} {Reviews of Modern Physics}\ }\textbf
  {\bibinfo {volume} {36}},\ \bibinfo {pages} {748} (\bibinfo {year}
  {1964})}\BibitemShut {NoStop}%
\bibitem [{\citenamefont {Friedrich}\ and\ \citenamefont
  {Haarer}(1984)}]{Friedrich1984}%
  \BibitemOpen
  \bibfield  {author} {\bibinfo {author} {\bibfnamefont {J.}~\bibnamefont
  {Friedrich}}\ and\ \bibinfo {author} {\bibfnamefont {D.}~\bibnamefont
  {Haarer}},\ }\href {\doibase https://doi.org/10.1002/anie.198401131}
  {\bibfield  {journal} {\bibinfo  {journal} {Angewandte Chemie International
  Edition in English}\ }\textbf {\bibinfo {volume} {23}},\ \bibinfo {pages}
  {113} (\bibinfo {year} {1984})},\ \Eprint
  {http://arxiv.org/abs/https://onlinelibrary.wiley.com/doi/pdf/10.1002/anie.198401131}
  {https://onlinelibrary.wiley.com/doi/pdf/10.1002/anie.198401131} \BibitemShut
  {NoStop}%
\bibitem [{\citenamefont {K\"onz}\ \emph {et~al.}(2003)\citenamefont {K\"onz},
  \citenamefont {Sun}, \citenamefont {Thiel}, \citenamefont {Cone},
  \citenamefont {Equall}, \citenamefont {Hutcheson},\ and\ \citenamefont
  {Macfarlane}}]{Konz2003}%
  \BibitemOpen
  \bibfield  {author} {\bibinfo {author} {\bibfnamefont {F.}~\bibnamefont
  {K\"onz}}, \bibinfo {author} {\bibfnamefont {Y.}~\bibnamefont {Sun}},
  \bibinfo {author} {\bibfnamefont {C.~W.}\ \bibnamefont {Thiel}}, \bibinfo
  {author} {\bibfnamefont {R.~L.}\ \bibnamefont {Cone}}, \bibinfo {author}
  {\bibfnamefont {R.~W.}\ \bibnamefont {Equall}}, \bibinfo {author}
  {\bibfnamefont {R.~L.}\ \bibnamefont {Hutcheson}}, \ and\ \bibinfo {author}
  {\bibfnamefont {R.~M.}\ \bibnamefont {Macfarlane}},\ }\href {\doibase
  10.1103/PhysRevB.68.085109} {\bibfield  {journal} {\bibinfo  {journal} {Phys.
  Rev. B}\ }\textbf {\bibinfo {volume} {68}},\ \bibinfo {pages} {085109}
  (\bibinfo {year} {2003})}\BibitemShut {NoStop}%
\end{thebibliography}%

\end{document}